# *Hierarchy, dimension, attractor and self-organization — dynamics of mode-locked fiber lasers*

Huai Wei[1]*, Bin Li[2,4], Wei Shi[3]*, Xiushan Zhu[4], Robert A. Norwood[4], Nasser Peyghambarian[4], and Shuisheng Jian[1]

[1]Institute of Light Wave Technology, Beijing Jiaotong University, Beijing 100044, China

[2]Department of Communication Engineering，Communication University of China，Beijing 100024, China

[3]College of Precision Instrument and Optoelectronics Engineering, Tianjin University, Tianjin 300070, China

[4]College of Optical Sciences, University of Arizona, Tucson, Arizona 85721, USA

Corresponding author: Huai Wei (hwei@bjtu.edu.cn), Wei Shi (shiwei@tju.edu.cn)

**Abstract:** Mode-locked fiber lasers are one of the most important sources of ultra-short pulses. However, A unified description for the rich variety of states and the driving forces behind the complex and diverse nonlinear behavior of mode-locked fiber lasers have yet to be developed. Here we present a comprehensive theoretical framework based upon complexity science, thereby offering a fundamentally new way of thinking about the behavior of mode-locked fiber lasers. This hierarchically structured frame work provide a model with and changeable variable dimensionality resulting in a simple and elegant view, with which numerous complex states can be described systematically. The existence of a set of new mode-locked fiber laser states is proposed for the first time. Moreover, research into the attractors' basins reveals the origin of stochasticity, hysteresis and multistability in these systems. These findings pave the way for dynamics analysis and new system designs of mode-locked fiber lasers. The paradigm will have a wide range of potential applications in diverse research fields.

**Introduction:**

As ideal ultra-short pulse sources, mode-locked lasers, especially fiber based mode-locked lasers, have generated great interest because of their inherent advantages and attractive properties [1][2][3]. The interplay among the many factors (nonlinear, dispersion, positive and negative feedback) in the cavity give the pulses in mode-locked fiber lasers rich and complex nonlinear dynamics. Typical pulse shapes include $sech^2$[2], public (self similar)[4-6] and flat top DSR (dissipative soliton resonance[7-11]) etc. The behavior of the pulses can be single pulse[2], multi-pulses[12-15], Q-switched mode-locking[1], and unstable pulses with periodic or non-periodic fluctuation etc. [16-21]. What is more, stochastic phenomena, hysteresis and multistability [22-25] further reflect the complexity of the mode-locked fiber lasers' dynamics. The broad range of temporal scales (from the femtosecond scale for pulse detail to the millisecond scale for Q-switched envelope fluctuations) presents significant difficulties to the analysis of mode locked lasers. There are long-standing problems with respect to the description, understanding and control of the complex dynamics governing the behavior of mode-locked fiber lasers. Extensive effort has been devoted to the understanding of mode-locked fiber lasers; fundamental equations were developed [2],[26-28], new kinds of pulses were discovered [4-11], and multi-pulse phenomena were analyzed [12-15]. However, there is still a lack of an ideal framework with convenient mathematical representations and clear physical meaning to give a unified description for the various states of a mode-locked laser and determine the driving force behind the complex nonlinear behaviors observed. A theory which can offer a way of seeing order emerge from disorder is desired. This is required if we are to transform the random, erratic and unpredictable behavior of mode-locked fiber lasers into graceful patterns analogous to well-known and fascinating nonlinear phenomena (i.e., Bernard convection, von Kármán vortex streets, Belousov-Zhabotinsky reaction etc. [29-33]).

From a methodological point of view, most current models try to study complex nonlinear phenomena from the traditional reductionism approach. There is no denying that many important and useful results have been derived from these models by this approach. However, explaining all complex nonlinear phenomena with this approach is inefficient and impractical. More importantly, it is difficult to develop clear physical insight, especially for macroscopic phenomena emerging from highly structured complex nonlinear behavior.

Complexity is becoming a powerful approach to a wide range of problems in chemistry, biology, economics and geomorphology etc. [34-39]. Hierarchy and the multi-scale method provide useful guidelines for how to understand complex phenomena[34,38,39]. These methods go beyond reductionism, showing the benefits of "stepping back from the wall" to gain a broader and more meaningful view of a picture, so that we see that it is more than just a series of crazy swirls [34]. Here we make use of this methodology to analyze the dynamics of mode-locked fiber lasers.

In this paper, with the help of hierarchy and the multi-scale method, the problem can be simplified from an infinite dimensional problem to an iterative mapping with finite but changeable dimensionality. We can thus represent and analyze mode-locked laser systems at a more profound level. The complex dynamical behaviors of mode-locked lasers actually are the manifestations of various attractors with different dimensions under various conditions. The common existence of strange attractors in mode-locked laser system is predicted for the first time. We can see that the "messy traces" made by mode-locked laser's complex action are, in fact, "footprints of an elegant dance."

What is more, within this theoretical framework we can make an in-depth study of the origin of many nonlinear behaviors such as stochastic phenomena, hysteresis and multistability. Multi-attractors, fluctuations, and attractor basins' variation with control parameters are the core factors. The important roles they play in these nonlinear phenomena are clearly revealed.

Outside the field of laser technology, the methodology we use here will be also helpful in analyzing complex phenomena in many other nonlinear domains which contain dissipative solitons as general concepts (e.g. condensed matter physics, fluid mechanics etc.).

**1. Hierarchical structure, the coarse grain method and the multi-dimensional model**

The mode-locked laser is an infinite dimensional dynamical system with feedback structure (see Fig.1 a.), evidencing a nonlinear system with spatio-temporal complexity [29][30]. What is more, the phenomena we are concerned about have a broad range of temporal scales. An all-inclusive model that addresses all details of the system is inefficient and has serious practical difficulties in implementation.

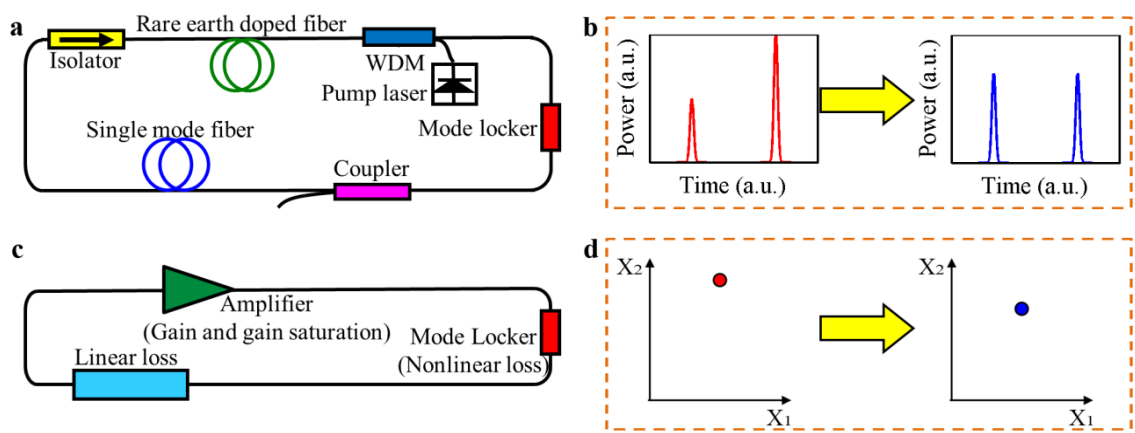

**Figure 1.** Schematic diagram of mode locked laser

Large composite systems, despite their complexity on the small scale, sometimes crystallize into large-scale patterns that can be conceptualized rather simply [34]. When focused on these macroscopic phenomena, we can divide the complex system into several levels. By using the hierarchical and coarse grain method, we can see that nonlinear dynamical phenomena emerge at the macro level, which allows key aspects of the system to be separate from extraneous details [34], [37-40]. The mode-locked laser is such a system with a hierarchical structure that can be naturally divided into three levels. (see Fig.2.).

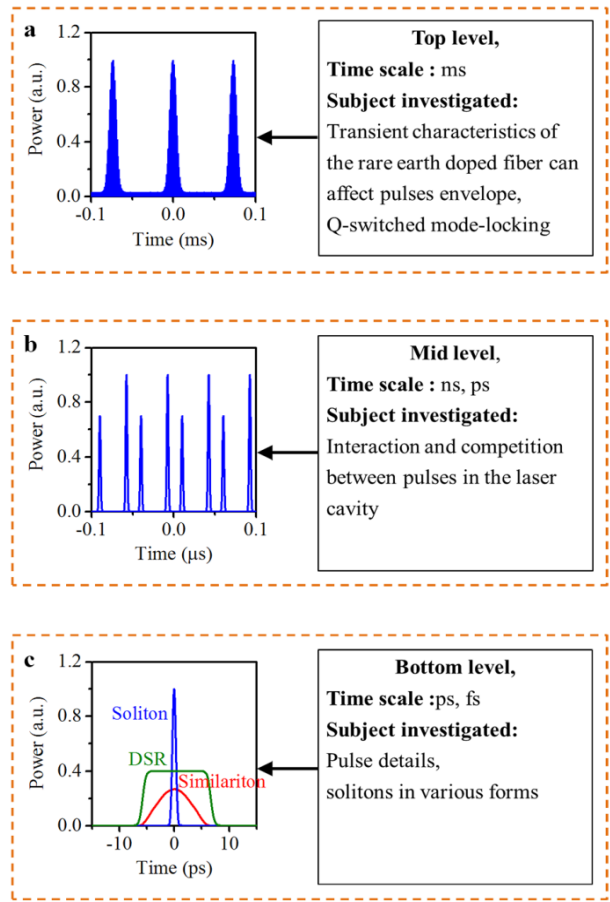

**Figure 2.** Hierarchy modeling methodology for dynamics of mode lock laser. Variables characterizing the system dynamics are arranged by temporal scale

At the bottom level of the description exist the pulse details, the dispersion and nonlinear dispersion and nonlinearity determine the characteristics of the pulses; these pulses are general dissipative solitons of various forms [7]. These solitons commonly have stable shapes in practice. At the middle level, gain and saturation effects result in interaction and competition between pulses. These factors allow the pulses to evolve toward a variety of well-organized complex patterns. At the top level, the transient characteristics of the rare earth doped fiber can affect the pulse envelope and facilitate the appearance of Q-switched mode-locking.

The pulse details can be deduced from the governing partial differential equations (Ginzburg-Landau equation) [2], [27], which correspond to a continuous infinite dimensional dynamical systems. There are many reports on models of the pulse details in the literature (e.g. [7] is a good review article), and thus analysis of this level will not be addressed herein due to space limitations. As a bridge that connects micro and macro phenomena, the intermediate-level is crucial part of the hierarchical model, and will be our primary focus. The top-level view and interactions among the various levels will be discussed concisely near the end of this paper.

The pulses in a mode-locked laser exhibit quantization [12, 13], which allows the description for the mid temporal scale level to be simplified. We ignore the details of the pulse, i.e., a pulse is considered to be a basic unit (a grain). Existing models [15] give a geometrical description of the onset of multi-pulsing in mode-locked laser cavities, and as such is an embryonic coarse grain model. We have improved upon this description and incorporate at it within a hierarchical model. Next, in a more important step, we employ "dimensions" and "phase space" concepts to provide a deeper analysis, as will fully emerge below. The pulse energy $E_j$ (the subscript "j" is the number index of the pulse in the laser cavity) is determined by the equivalent width $t_j$ (pulse duration) and equivalent amplitude $x_j$ (Equation (1)). In the simplest case, pulse energy is independent of pulse duration, and the picture reduces to the model based on pulse energy as in [15]. In the more complex case, if pulse amplitude is relevant to pulse energy, Equation (2) should be used. Gain saturation has effects on the laser's average power as determined by the pulse number in the laser cavity and the individual pulse energies (Equation (3, 4)). In addition, nonlinear loss coming from the mode locker effects every pulse, respectively. There are many kinds of models can be used for the effect of the mode locker (f(x) in Equation (5) ) [15,41-42].

$$E_j = \int p_j dt = x_j t_j = x_j t_{eff} \quad (1)$$

$t_{eff} = T(p)$ \quad (2)

$$G = e^g \quad (3)$$

$$g = \frac{g_0}{1 + \frac{\sum_{j=1}^{m} E_j}{E_{sat}}} \qquad (4)$$

$$L_j = f(x_j) \qquad (5)$$

The coarse grain method transforms an infinite dimensional dynamical system(Fig. 1a,b) into a discrete dynamical system generated by an iterated map with countable dimensions(Fig. 1c,d). The phase space for the mode-locked laser system is an $R^m$ space, where a vector in $R^m$ space can be used to represent a state at a certain moment in time in a mode-locked laser system. Every component of the m-dimensional vector is a representation of the equivalent power of each pulse. The dimension "m" corresponds to the number of pulses in the laser cavity.

The intracavity pulse number, a primary focus, reflects the competition and influence among the pulses, which can be described and analyzed by a high dimensional discrete dynamical system. The time dependent states of the pulses can be expressed as a trajectory in the phase space. We need to note that the "m" in the m-dimensional $R^m$ space is neither infinite nor invariant, as will become clear below. Changes in the laser's parameters are often accompanied by extension or collapse of the dimensions in phase space. Mathematically, this is a mapping with changeable dimensionality.

**2. Attractors, the existence states of pulses in mode-locked lasers**

States of pulse in mode-locked lasers have the characteristics of complexity and diversity. Describing and explaining these complex states and the dynamics of conversion between these states are the core topics of interest. The dynamical system concepts of "dimension", "attractor" and "attraction basin" are key to establishing adequate descriptions of the mode-locked laser behavior.

As previously mentioned，the laser state can be represented by a vector in $R^m$ space, effectively reducing an infinite dimensional problem to an m-dimensional one. This can be thought of as taking a generalized Poincaré section in the infinite dimensional space. The vector can also function as an "order parameter" in synergetics[43,44], an approach to describing system phenomena. In this case, the abstracted model for the mode-locked laser system at the mid temporal scale level of current interest becomes an iterative map. The trajectory of this iterative map is a series of discrete points in multi-dimensional space that describe the evolution of the laser's state. After a sufficient number of iterations, the system settles down to its final state. The attractors (or stable fixed points) of the dynamic system emerge.

The attractors give a clear mathematical description for the states of the mode-locked laser at mid-level of the hierarchy, providing us with the required perspective to obtain a global image of the dynamics of the mode-locked laser. All the possible laser states can be found in this image (Table.1).

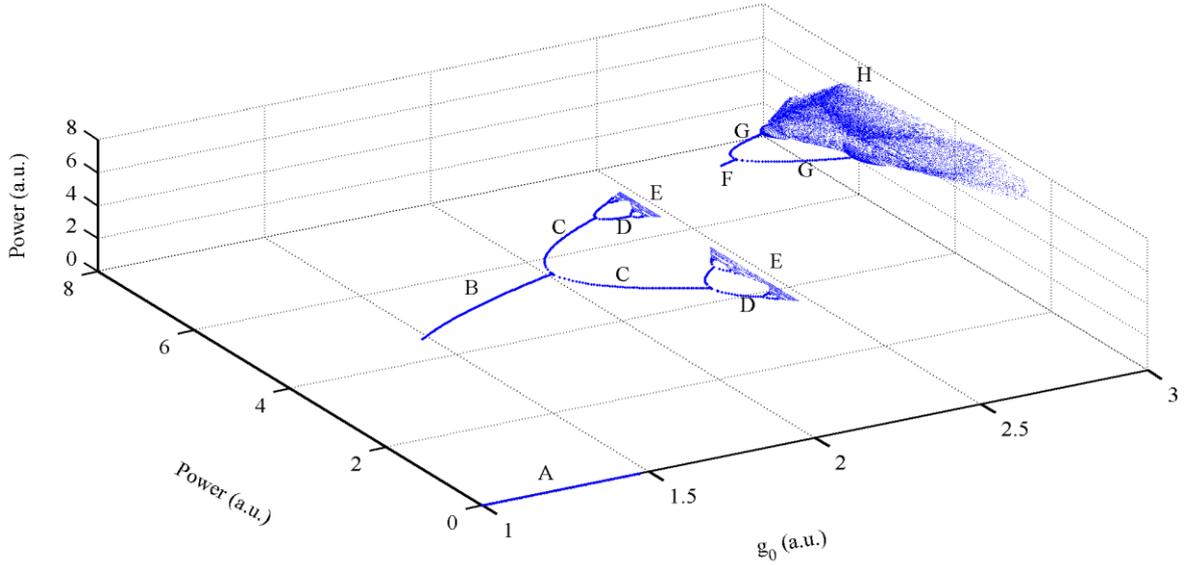

**Figure 3.** Bifurcation orbit diagram in multi-dimensional phase space under certain initial conditions with different gain coefficient $g_0$.
Point A ($1<g_0<1.46$), there is no pulse; point B ($1.46<g_0<2.02$), stable single pulse state; point C ($2.02<g_0<2.32$) and D ($2.32<g_0<2.40$), single pulse with periodic fluctuations; point E ($2.43<g_0<2.47$), single pulse that exhibits chaotic fluctuation; point F ($2.47<g_0<2.53$) stable double pulse state; point G ($2.53<g_0<2.74$), double pulse with periodic fluctuations; point H ($g_0>2.82$), chaotic fluctuation.

Fig. 3 show a bifurcation diagram in multi-dimensional phase space under certain initial conditions with different gain coefficient $g_0$. In Fig. 3 the phase space is shown as a vertical plane. The third dimension represents the control parameter - - gain coefficient $g_0$. The composite space of the phase space and the control parameter is the bifurcation diagram illustrating the whole complex dynamics.

The attractor for a given gain coefficient $g_0$ is the cross-section at $g_0$ perpendicular to $g_0$-axis. The attractors have diverse and varied features. From Fig. 3 we can see that the working state of a mode-locked laser changes according to the value of the parameter $g_0$. Different attractors mean that the dynamical system can have different patterns of behavior. Fixed point, periodic orbit and chaos attractors represent stable pulses, pulses with periodic fluctuation and chaotic fluctuation, respectively.

Compared to common nonlinear systems, mode-locked lasers show more interesting unique characteristics based on their dimensionality. Attractor shape and topological structure are not the only origins of attractor diversity, as dimensionality also plays an important role. Let's investigate Fig. 3 for an example. We see that when $g_0$ is too low ($g_0<1.46$), there is no pulse (Fig. 3, point A). Under appropriate conditions for the gain coefficient ($1.46<g_0<2.02$), the laser operates stably in the single pulse state, and the attractor is a fixed point in phase space (Fig. 3, point B). At points C ($2.02<g_0<2.32$) and D in Fig. 3 ($2.32<g_0<2.40$), the attractor becomes two (or more) discrete points. This means that the operating state is a single pulse with periodic fluctuations. At point E ($2.43<g_0<2.47$) in Fig. 3, the attractor is a set of discrete points, i.e. a strange attractor in 1 dimension. This means that there is a single pulse that exhibits chaotic fluctuation. At point F ($2.47<g_0<2.53$), the attractor is once again a fixed point. But different from point B, point F is a point in 2 dimensional space. At points G ($2.53<g_0<2.74$) and H in Fig. 3 ($g_0>2.82$), periodic orbits and chaos attractors emerge in 2D phase space.

In fact, when increasing the gain coefficient to higher values, we can see attractors in higher dimensional space. But in these cases the bifurcation diagram become a four (or even higher) dimensional graph, to which it is difficult to give an intuitive graphical representation.

Even these simple examples can provide many physical insights regarding the dynamics of mode-locked lasers. An attractor is generally a set of points in the phase space, so for a given parameter $g_0$ and a phase space with large enough dimensionality (as initial condition), all the points of the attractor will finally be located in a sub space with limited dimensionality. For different values of the parameter $g_0$, the dimensionality of the sub-space may be different. In other words, the variation of the gain coefficient can cause extension or collapse of the dimensions in phase space increasing dimensionality can provide space for increasing complexity. In this process, fluctuations seed the creation of new dimensions, and the control parameter (gain coefficient) determines the dimensionality of the attractor.

By using the coarse grain method we see that the behavior of complex systems can be successfully modeled by means of a few macroscopic quantities. Such macroscopic observables play the role of a thermodynamic order parameter in the synergetic framework [43,44]. Here the phase point in the multidimensional mapping phase space with variable dimension represents the state of mode-lock laser. Switching of laser's operating states by changing the control parameter can be interpreted as order parameter switching between attractors in the phase space. Dimension mutation of this parameter (extension or collapse of the dimensions) occurs at the bifurcation points on the bifurcation diagram, essentially acting like a phase transition process (nonequilibrium phase transition). We will see that variation of the attractor basin is the driving force for this transition process and the random fluctuation acts to seed the transition.

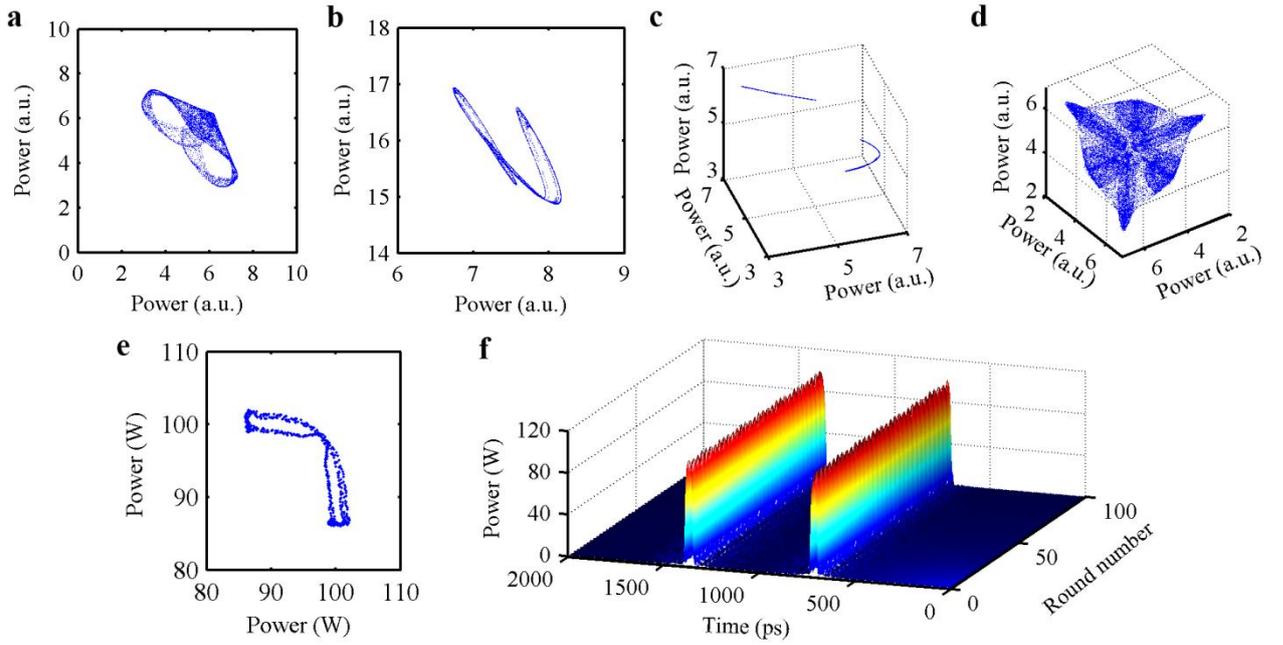

**Figure 4.** Chaos attractor in phase space. **a**, **b** shows the attractor for two pulses. **c**, **d** shows the attractor for three pulses. **e f** shows the result derived by full numerical model

Complex and diverse operating states can be presented by attractors with various styles and shapes. The strange attractors exist in a variety of forms as seen in Fig. 4. Table. 1 summarizes and lists the operating states of a mode-locked laser. It is noteworthy that, by using the model and methodology we provided, we have identified a new state for pulses in mode-locked lasers which has not yet reported, namely coexisting multi-pulses that form a chaos attractor in phase space. Fig. 4 shows the attractors for two pulses and three pulses. The Lyapunov exponent (Figure S1) and correlation dimension [45-47] (Figure S2, Figure S3 ) confirm the existence of strange attractors.

The existence of the chaos attractor is also confirmed by using a traditional direct numerical model. Fig. 4e,f show the pulse in the time domain and Poincaré sections of the pulses (taken at the peak power of pulses) in phase space derived by direct numerical simulation.

**3. Attraction basin, the mechanism behind stochastic phenomena, hysteresis and multistability**

In the previous section, we have analyzed operating states of the mode-locked laser. In this section, we will discuss another important problem: the dynamics of conversion between these states, including state transition and pulse start-up. There are many complex and puzzling phenomena (stochastic phenomena, hysteresis and multistability etc.) involved in these dynamics [22-25]. However, the mechanisms behind many of these phenomena are still obscure and present another opportunity for our model.

For a thorough grasp of the state transition problem we need a global understanding of the laser behavior under all possible initial conditions. Direct numerical modeling [48] would require an enormous amount of computation and suffer from many uncontrollable factors (e.g. pulse splitting, pulse deformation and ASE noise). These factors make a global quantitative analysis for the dynamics of the system challenging. We have seen how the coarse grain model for the mid-temporal scale level reduced the infinite dimensional problem in real space to a phase space problem with limited dimension. This approach can provide improved controllability, reduce variations from uncertainties, and enable quantitative analysis of the key features.

Tracking the evolution of all points in phase space under a given set of parameters mathematically corresponds to the attraction basin under different control parameters. Attraction basin phase portraits provide an excellent perspective, providing a glimpse of the whole picture and clues to the origin of little understood phenomena in mode-locked lasers.

Fig. 5 and Figure S4 show the attractors and attractor-basin phase portraits for different gain coefficients ($g_0$). By analyzing this series of pictures we can recognize patterns and then identify the source of stochastic phenomena, hysteresis and multistability. In Fig. 5a – 5b, the gain coefficient ($g_0$) is 1.70. There are 2 kinds of attractors: 1) fixed point at the coordinate origin which means no pulse; 2) fixed point on the x-axis or fixed point on the y-axis. In both of the latter cases a stable single pulse exists in the laser cavity; these cases are indistinguishable at the macro level. Fig. 5a shows the corresponding attractor-basin where the green area in Fig. 5a shows the attractor-basin for an attractor at coordinate origin. If the initial state points fall in this region, the laser is unable to sustain a pulse. The blue and red areas are, respectively, the attractor-basins for the other attractors. They have axially symmetric shapes about the symmetry axis y=x. In this case the laser can only support one pulse; the pulse with highest initial power can survive in the competition with other pulses.

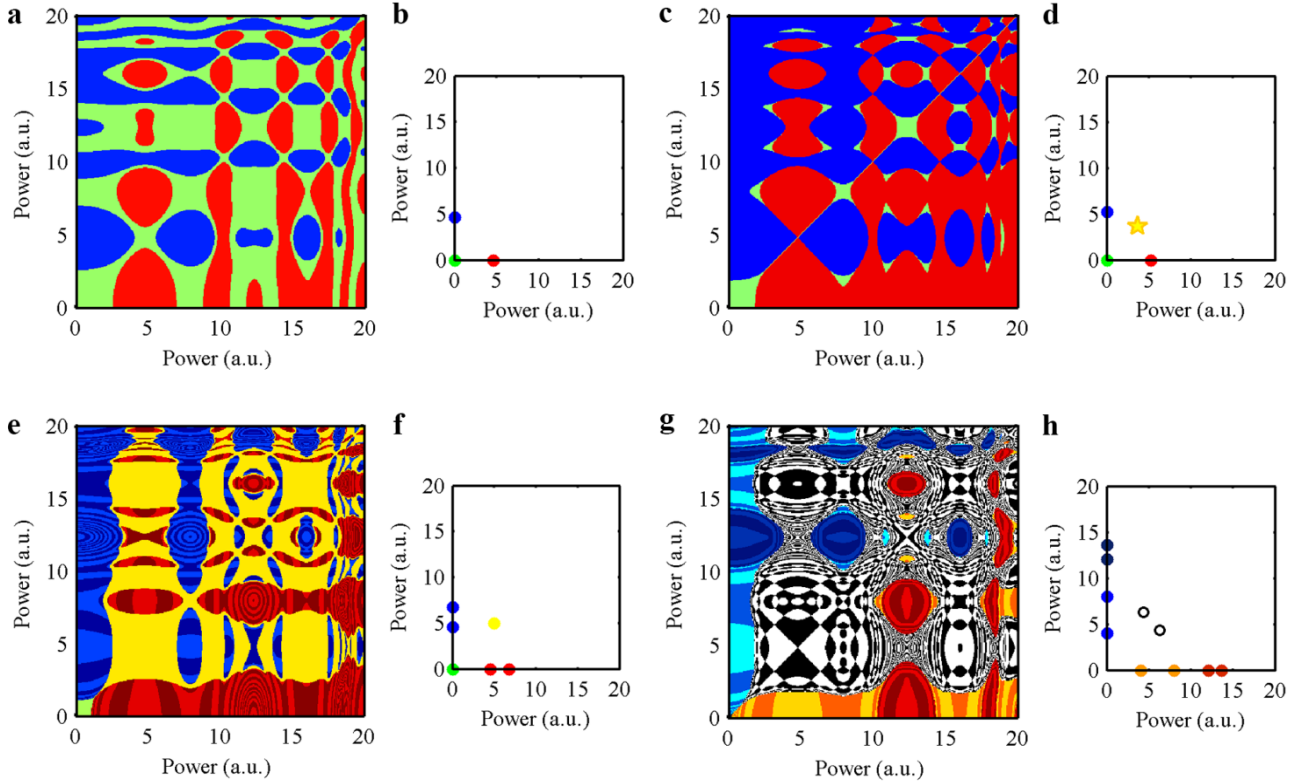

**Figure 5.** The attractors and attractor-basin phase portraits for different gain coefficients.

Any initial condition is a point in phase space. A square region for possible initial condition in phase space is subdivided into 500×500 cells. Do the iteration and tracking the points on the grid until get the attractors. Then we can derive the attractor-basin. Different colors correspond to different attractors. **a, c, e** and **g are** attractor-basin phase portraits for different gain coefficients ($g_0$);    **b, d, f** and **h** are attractors for **a, c, e** and **g**. (**The parameters for Fig. 5 in detail please see method**)

For Fig 5. **a** and **b**, $g_0$=1.7. There are three attractors in **b** (blue, red and green points). Blue point and red point are stable single pulse state. Green point means no pulse. The attractor-basins for the attractors are shown in **a** (use the same color as the corresponding attractor in **b** )

For Fig 5. **c** and **d**, $g_0$=2.0. There are four attractors in **d** (blue, red, green and yellow points). Blue point and red point are stable single pulse state. Green point means no pulse. A new attractor (the point with yellow star marker) means that the laser can support 2 pulses under a narrow range of initial conditions.

For Fig 5. **e** and **f**, $g_0$=2.5; There are four attractors in **f** (blue, red, green and yellow points). Blue points and red points are single pulse in periodic fluctuation state. Green point means no pulse. Yellow point means stable double pulses.

For Fig 5. **g** and **h**, $g_0$=3.0; There are three attractors in **h** (blue & dark blue, orange & red, black circle). Blue & dark blue points and orange & red points are single pulse in periodic fluctuation state. Black circle is for double pulses in periodic fluctuation state.

Fig. 5c is similar to Fig. 5a; however, some changes occurred in the basins of attraction as a result of different gain coefficient; note that the green areas significantly reduced, while the red and blue area become closer to one another. We can see the emergence of a very small yellow area, which is the attractor-basin for a new attractor (the point with star marker in Fig. 5d). This means that the laser can support 2 pulses under a narrow range of initial conditions.

In Fig. 5e, f the gain coefficient ($g_0$) is 2.5. Compared with Fig. 5c, d, fixed point attractors on x-axis and y-axis become a periodic orbit which represents a single pulse with periodic fluctuations. Smaller green areas and larger yellow areas indicate that the laser is easer to start up and more likely to enter the double pulse state than the laser of Fig. 5c, d.

In Fig. 5g, h (and Figure S4a, b) the gain coefficient ($g_0$) is 3.0. Periodic fluctuations appear in both the single pulse and double pulse states. The area for the fixed point attractor basin at the origin has now almost disappeared. This means that the laser will have good self-starting characteristics. Finally, in Figure S4c, d, $g_0$=4.5 and the attractors become chaotic attractors. Power fluctuations are very large under this condition.

There have been many reports that describe the nonlinear behavior in the laser state conversion process (stochastic behavior, hysteresis and multistability) [22-25]. We now show that we can look beyond the surface to understand the underlying mechanisms of these well-known nonlinear phenomena by performing an in-depth study of the attractor-basin phase portraits.

Stochastic phenomena in mode locked lasers means that the same system under the same control parameters can enter different states stochastically. The pattern of the attractor-basin indicates the relationship between the system's initial state and final state. Commonly, there are some coexisting attractors as described above. In this case, even under the same control parameter, the system will enter a different final state if the system has an initial value that belongs to a different attractor-basin. So uncertainty in the initial state can cause the system to exhibit stochastic characteristics. For the mode-locked fiber laser system, stochastic phenomena commonly emerge in three cases: the startup process, the pulse splitting process, and the parameter switching process. In the first two cases, ASE noise and unstable broken pulses give the system a stochastic initial position in some region of the phase space. In the third case, the system starts in either a periodic orbit or a strange attractor. The position of the point representing the state of the system changes with time in phase space. Its trajectory may cross multiple attractor-basins for the new parameter value which the system will be switched to. Then the state that the laser will finally settle down to is determined by two factors: the initial state at the moment the system was switched to the new parameter value, and the attractor-basins for the new parameter value.

Thus, if we know the statistical properties of the initial conditions (statistical properties of the initial phase point in the phase space) and distribution of the system's attractor basin, we can determine the probability of a certain operating state into which the system will settle.

The coexistence of multi-attractors and their attractor-basins is also the origin of hysteresis and multistability. Let's illustrate this using Fig. 5 as an example. We change the system status by adjusting gain coefficient $g_0$ as a control parameter. In order to make the laser self start we need to increase $g_0$ to a high level. From Fig. 5 we can see that increasing $g_0$ can cause expansion of the attractor-basins for single pulse （and double pulse）status and the shrinking of the attractor-basins for 0 pulse. It can enable the system enter single pulse state (red and blue area in Fig.5 g) from a low power noise initial state. If $g_0$ is increased to an even higher level, we can increase the area of the attractor-basins for double pulse status and get 2 pulses in the laser cavity (Figure.S4). To change the double pulse state to a single pulse state, we should decrease $g_0$ to reduce the attractor-basins for double pulses (Fig. 5c). If we want to extinguish the pulse in the laser, $g_0$ should be decreased to a lower level than that of Fig. 5a. This hysteresis and multistability process is shown in Fig.6. The high gain at laser start up and the decreased gain to obtain a stable single pulse is generally observed for mode-locked fiber laser and has been verified here within the current framework.

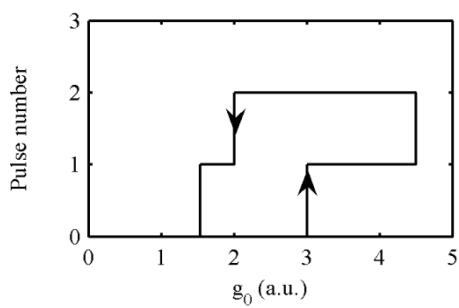

**Figure 6.** Hysteresis and multistability for pulse number in mode locked laser.

We see that the variation of the attractor-basins with the control parameter plays a crucial role in the onset of hysteresis and multistability. The mechanics is as follows: attractors locate in attractor-basins, and fluctuation makes the phase point which corresponds to the system state move stochastically in the neighborhood of the attractor in phase space. The conversion between the system's states is represented by the switching between different attractors. In this course of this switching, the phase point needs to exit the current attractors-basin，and jump (drop) into a new attractors-basin. In order for the system to escape from an attractors-basin we can change the pattern of attractor-basins by adjusting the control parameter. Shrinking or even vanishing of the basin of the current attractor together with the expansion of the adjacent attractor-basin can induce state conversion of the laser. When the conversion is complete, the system phase point must enter a new attractor-basin with a large area. If we want to change the system state again, we can adjust the control parameter to decrease the area of this new attractor-basin. On the macro level, the accumulation of control parameter changes results in a sudden change in the system state. To predict the system's future state, not only the control parameter but also the history of the system must be known.

Briefly speaking, the critical value of the state transition is jointly determined by the magnitude of the fluctuations as well as the shape and area of the attractor basin. Decreases to the local area of the attractor basin and increases in the fluctuations facilitate the state transition. The coexistence of fluctuations and the multi attractors is the fundamental origin of multistability and hysteresis. In practical applications, designing a laser system's attractor basin to have a large area and high stability is the way to get a stable high power single pulse mode locking laser. In addition, the study of the phenomenon of hysteresis and multistability has potential importance for optical storage.

## 4. Outlook and discussions

We have proposed the multi-level hierarchy model for mode-locked lasers and discussed it in detail for the mid-scale level (Fig.2). It should be noted that for the actual system, there are interactions between the various levels, where these interactions constitute the coupling channels between levels. Due to limited space in this paper, we will only point out the sources of these coupling channels without an in-depth discussion. These include the following:

1. the character of the "grain" in mid-level, i.e., the relationship between the effective peak power and the effective pulse duration (for example DSR the pulse duration increases while the amplitude remains almost unchanged);
2. the threshold of pulse splitting and the property of the new pulses that emerge after pulse splitting;
3. large time scale processes such as Q-switching, in which case the transient characteristics of the rare-earth doped fiber should be taken into account. Statistical properties and pulse envelopes can be used for simplification.

In addition, it needs to be pointed out that for the model discussed in this article, the pulses should are assumed to have some special and stable shapes (e.g., generalized solitons). For pulse waveforms without a stable shape (e.g. rogue waves[49]), a separate discussion is required.

From the point of view of dissipative system theory, the mode-locked fiber laser is a thermodynamically open system operating far from thermodynamic equilibrium in an environment with which it exchanges energy. The laser's excitation occurs when the energy supply exceeds the threshold. For pulsed excitation, the mode locker is the key component for symmetry breaking in time domain. The symmetry breaking together with the joint action of saturation and feedback cause the emergence of temporal and spatial patterns. Under the action of self organization we can see that a system can vary from spontaneous emission in the disordered state to a continuous wave laser with a highly ordered state in the frequency domain, to a pulsed laser with a highly ordered state in time domain (pulse with a specific shape), to a laser with multiple pulses that shows ordered and chaotic macro behaviors. Within a theoretical framework containing the concepts of hierarchy, coarse grain, dimension, attractors, and self organization, a highly structured temporal and spatial pattern emerges. Furthermore, the analysis of the attractors' basin tells us that the fluctuations, the coexistence of attractors, and the variation of attractors' basins are the driving force for the system's evolution and macroscopic nonlinear behavior.

We have proposed a new method that can help us see through chaos and understand the complexities of the behavior exhibited by mode-locked lasers. It is worth pointing out that, the method we used here provides an important example for the analysis of a nonlinear system. On the methodological level, it will have a wide range of applications in the fields of physics, nonlinear dynamics and complex systems.

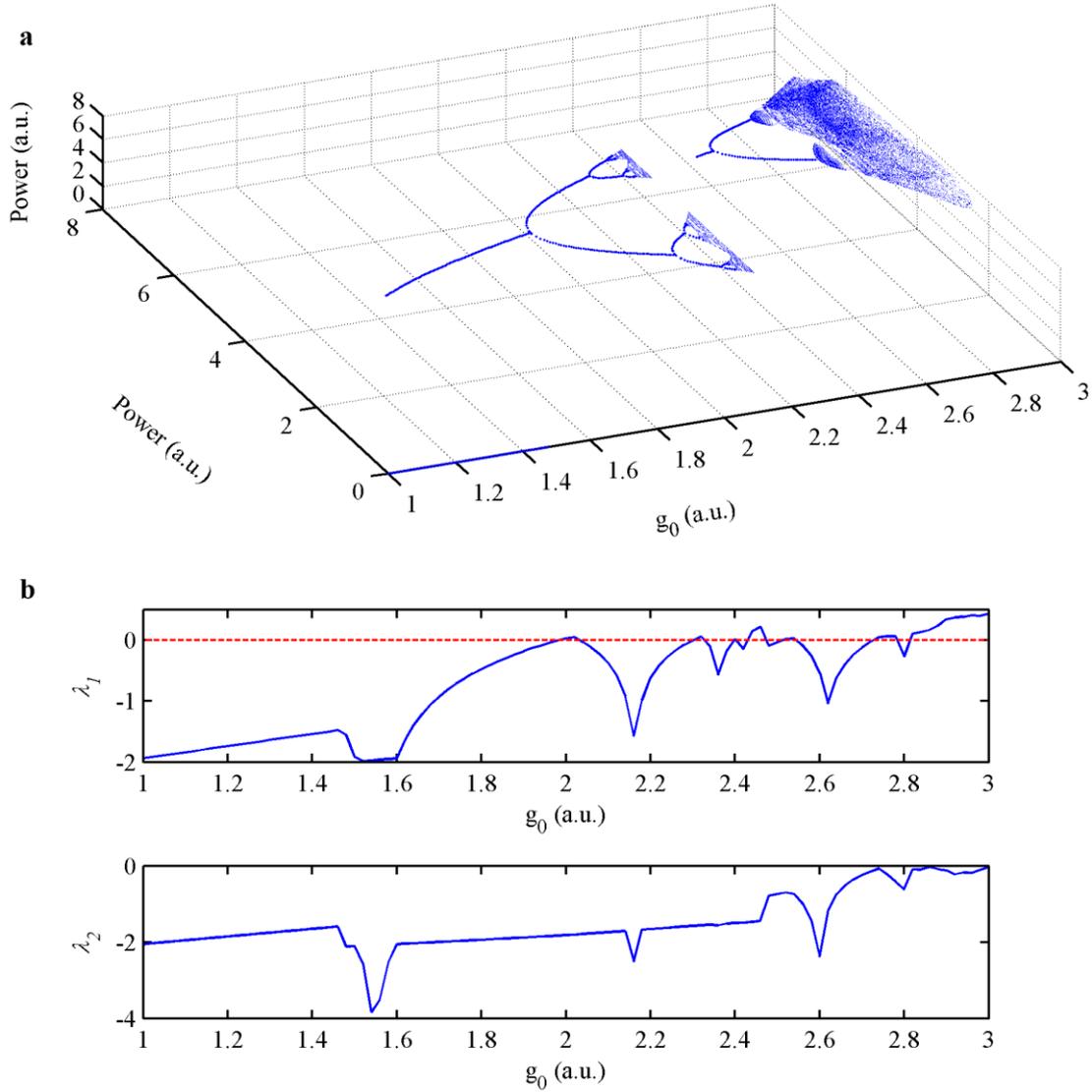

**Figure S1.** Bifurcation diagram and Lyapunov exponent

**a**, Bifurcation diagram in multi-dimensional phase space under certain initial conditions with different gain coefficient $g_0$.

**b** The first and second Lyapunov exponent (derived by Wolf's method)

**The parameters for Figure S1:**

$M_0=0.1$, $M_N=0.36$, $PM=8$, $P_\theta=0$ equation (M.1); $E_{sat}=5$ (Equation (5) in the main text).

The cavity loss caused by the coupler is 50% (50% of the power is extracted from the cavity)

The initial condition: $x_1=3.00$, $x_2=2.12$ (Equation (1) in the main text)

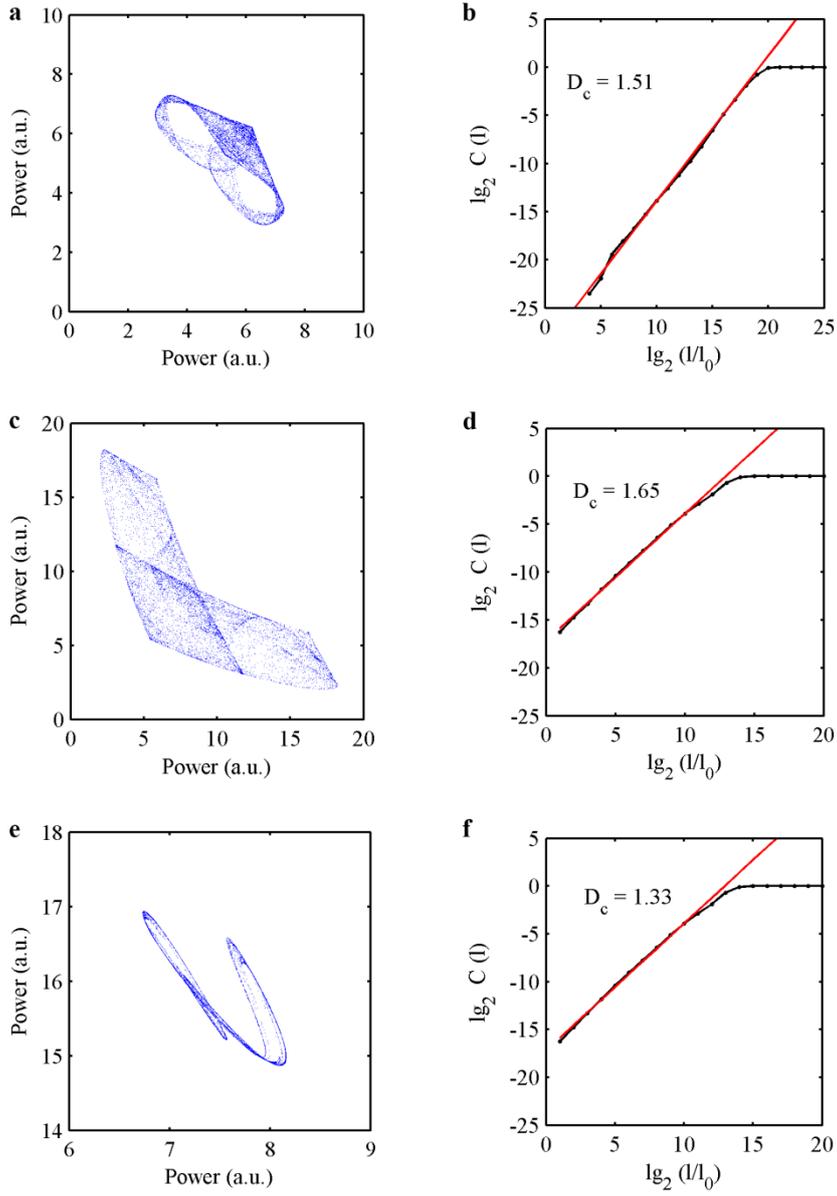

**Figure S2.** Strange attractors and correlation dimension for two pulses

$M_0$=0.1, $M_N$=0.3, PM=8 , $P_\theta$=0 equation (M.1);$E_{sat}$=5 (Equation (5) in the main text).

The cavity loss caused by the coupler is 50% (50% of the power is extracted from the cavity)

The initial condition : $x_1$=3.0 , $x_2$=2.12 (Equation (1)  in the main text )

The gain coefficient: $g_0$

$g_0$ =3.2 for **Figure S2 a,b**; $g_0$ =4.5 for **Figure S2 c,d**; $g_0$ =7.0 for **Figure S2 e,f**;

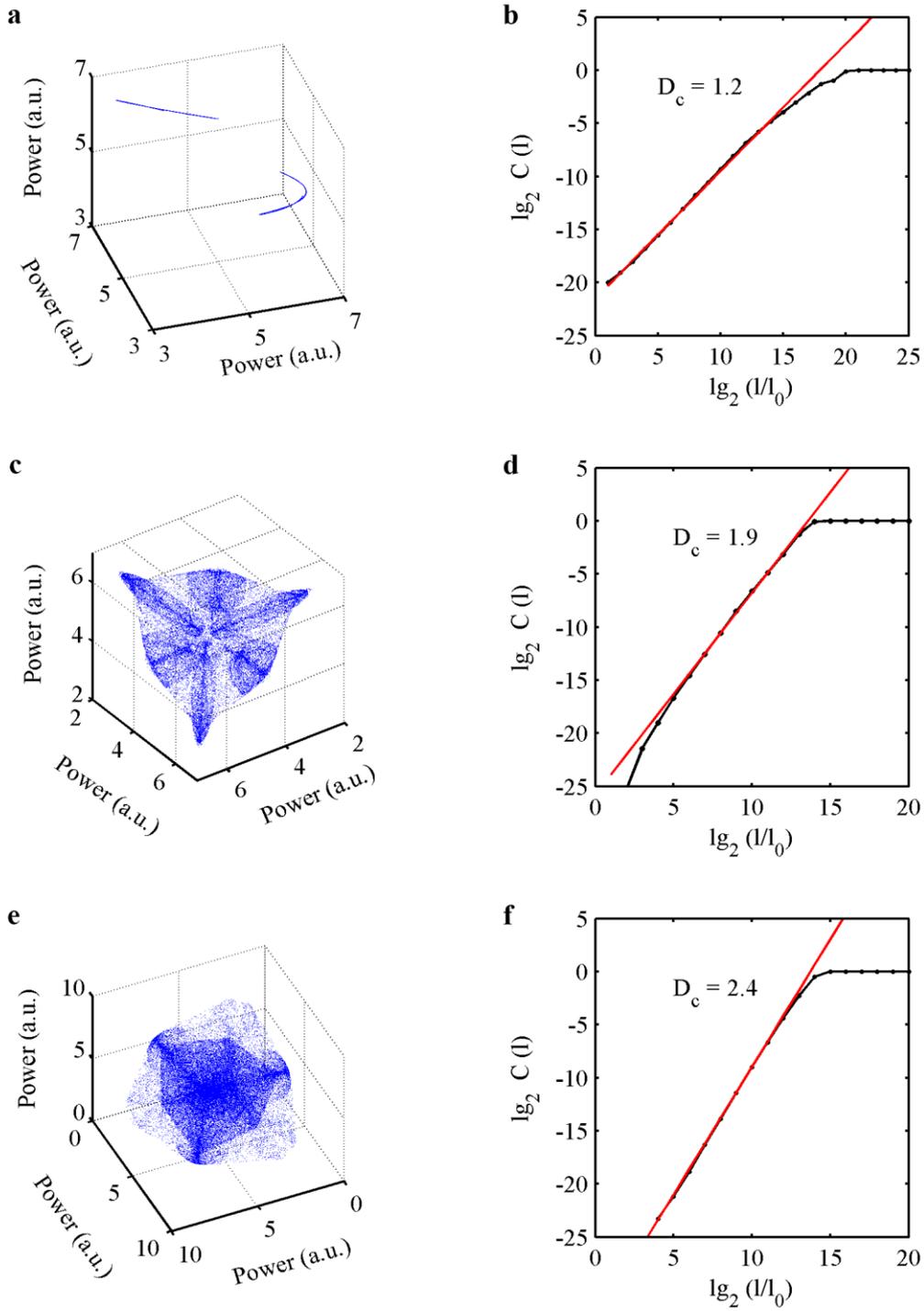

**Figure S3.** Strange attractors and correlation dimension for three pulses

$M_0$=0.1, $M_N$=0.3, PM=8 , $P_\theta$=0 equation (M.1);$E_{sat}$=5 (Equation (5) in the main text).
The cavity loss caused by the coupler is 50% (50% of the power is extracted from the cavity)
The initial condition : $x_1$=6.0 , $x_2$=7.12, $x_3$=8.0   (Equation (1)  in the main text )
The gain coefficient: $g_0$. $g_0$ =3.90 for **Figure.S3 a,b**; $g_0$ =3.95 for **Figure.S3 c,d**; $g_0$ =4.15 or **Figure.S3 e,f**;

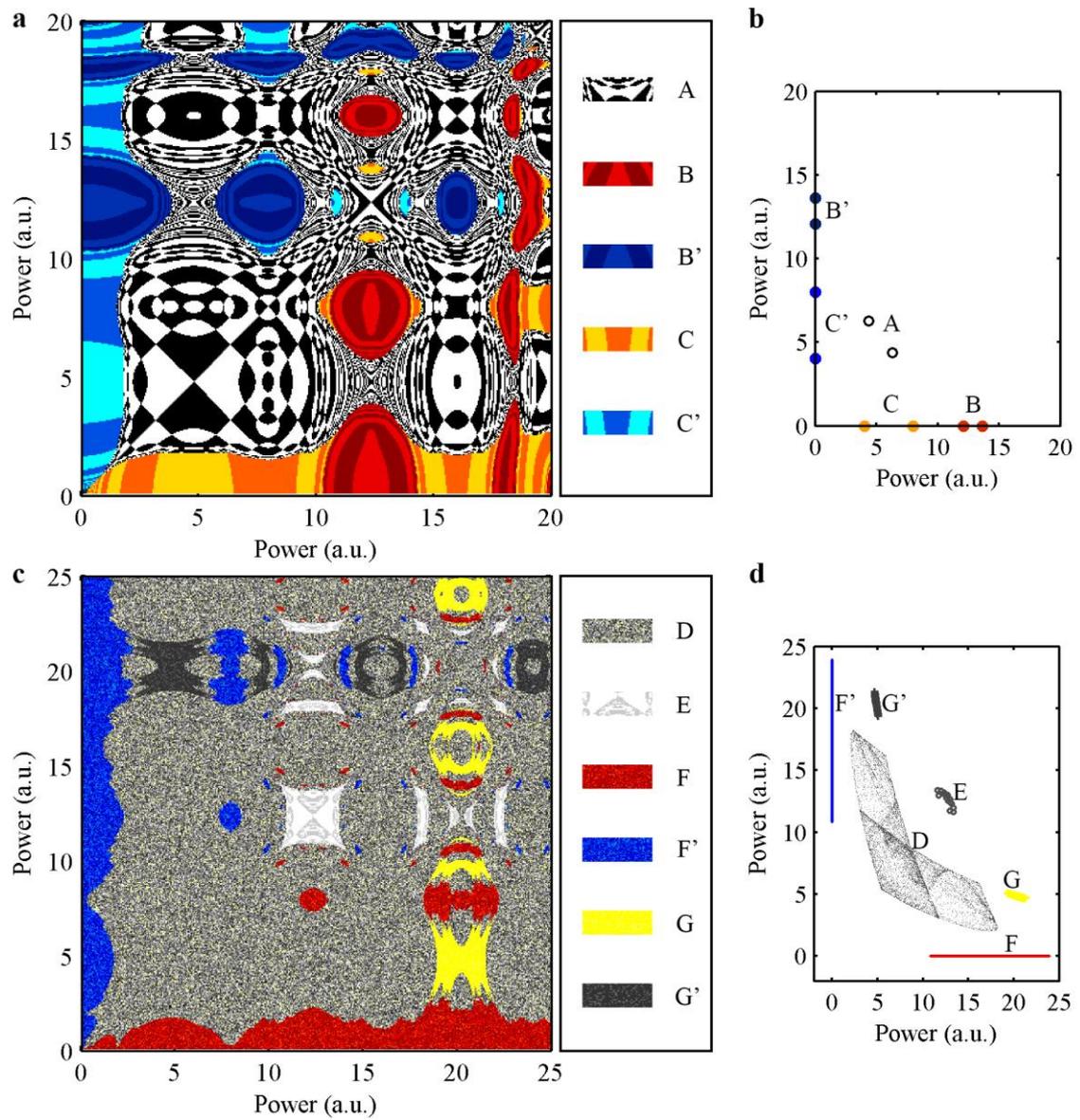

**Figure S4.** The attractors (**b, d**) and attractor-basin (**a,c**) phase portraits for different gain coefficient

$M_0$=0.1, $M_N$=0.3, PM=8 , $P_\theta$=0 equation (M.1);$E_{sat}$=5 (equation (5) in the main text).
The cavity loss caused by the coupler is 50% (50% of the power is extracted from the cavity).
The gain coefficient: $g_0$
For Fig S4. a b $g_0$=3.0; For Fig S4. c d $g_0$=4.5;

# Table.1

| Pulse number | Space dimension | Pulse state | The attractor and it's dimension | Figure |
|---|---|---|---|---|
| **Single pulse** | 1 | Stable single pulse | A point in 1D space; dimension: D=0 | Fig.3 point A |
| | 1 | Single pulse with periodic fluctuation | Two (or more) discrete points dimension: D=0 | Fig.3 point B,C |
| | 1 | Single pulse with chaos fluctuation | Strange attractor in 1D space dimension: 0<D<1 | Fig.3 point E |
| **Two pulses** | 2 | Stable two pulses | A point in 2D space; dimension: D=0 | Fig.3 point F |
| | 2 | Two pulse with periodic fluctuation | Two (or more) discrete points in 2D space dimension: D=0 | Fig.3 point G |
| | 2 | Two pulse with chaos fluctuation | Strange attractor in 2D space dimension :0<D<2 | Fig.3 point H, Figure S2 Fig.4 a,e,f  *1 |
| | 2 | Asymmetric two pulses | The attractor is asymmetrical about "x =y" | Fig.4 b, Figure S2.e *2 |
| **Three pulses** | 3 | ...*3 | ... | |
| | 3 | Three pulse with chaos fluctuation | Strange attractor in 3D space 0<D<3 | Figure.S3,Fig.4 c,d *4 |

***1, *2, *4 :** New states for pulses in mode-locked laser which have not yet reported

***3:** Similar to the previous case (Stable, periodic fluctuation, chaos fluctuation)

23